\documentclass[submission,copyright,creativecommons]{eptcs}
\providecommand{\event}{FROM 2023} % Name of the event you are submitting to

\usepackage{iftex}

\ifpdf
  \usepackage{underscore}         % Only needed if you use pdflatex.
  \usepackage[T1]{fontenc}        % Recommended with pdflatex
\else
  \usepackage{breakurl}           % Not needed if you use pdflatex only.
\fi

\usepackage{amssymb}
\usepackage{amsfonts}
\usepackage{amsxtra}
\usepackage{latexsym}
\usepackage{graphicx}
\newtheorem{theo}{Theorem}[section]
\newtheorem{theo*}{Theorem}
\newtheorem{lemma}[theo]{Lemma}
\newtheorem{definition}[theo]{Definition}
\newtheorem{remark}[theo]{Remark}
\newtheorem{cor}[theo]{Corollary}

\newcommand{\C}{{\mathbb{C}}}

\newcommand{\F}{{\mathbb{F}}}

\newcommand{\qed}{\hspace*{\fill}$\Box$}

\newcommand{\cell}[1]% #1 = text
{\makebox[1em]{#1}}

\title{Symmetric Functions over Finite Fields}
\author{Mihai Prunescu
\institute{Research Center for Logic, Optimization and Security (LOS), \\ Faculty of Mathematics and Computer Science, \\ University of Bucharest, Academiei 14, 010014 Bucharest, Romania}
\institute{and Simion Stoilow Institute of Mathematics of the Romanian Academy, 
\\ Research unit 5, P. O. Box 1-764, RO-014700 Bucharest, Romania.}
\email{mihai.prunescu@imar.ro, mihai.prunescu@gmail.com}
}
\def\titlerunning{Symmetric Functions over Finite Fields}
\def\authorrunning{Mihai Prunescu}
\begin{document}
\maketitle

\begin{abstract}
The number of linear independent algebraic relations among
elementary symmetric polynomial functions over finite fields is computed.
An algorithm able to find all such relations is described.
It is proved that the basis of the ideal of algebraic relations found by the algorithm 
consists of polynomials having coefficients in the prime field $\F_p$.

\thanks{A.M.S.-Classification: 14-04, 15A03.}
\end{abstract}

\section{Introduction}

The interpolation problem for symmetric functions over finite fields does not have a unique solution in terms of elementary symmetric polynomials. That is why it is useful to know more about the set of solutions, as sometimes we need a solution which is easier to express or faster to evaluate. The general situation will be described in the lines below. Some steps are detailed in the following sections.

Let $\F_q$ be a finite field of characteristic $p$, a prime number. Let $\F_q[X_1, \dots, X_n]$ be the $\F_q$-algebra of polynomials in variables $X_1, \dots, X_n$ over $\F_q$. The symmetric group $S_n$ acts on $\F_q[X_1, \dots, X_n]$ as a group of automorphisms of $\F_q$-algebras and the subalgebra $Q$ of fixed points is usually called the ring of symmetric polynomials. The algebra $Q$ is again a polynomial ring freely generated by the elementary symmetric polynomials \[e_i = \sum \limits _{1 \leq j_1 < j_2 < \dots < j_i \leq n} X_{j_1}\dots X_{j_i}\] for $i = 1, \dots , n$. So $Q = \F_q[e_1, \dots, e_n]$.

The vector space $Q$ has a basis given by the monomial symmetric functions $e_1^{k_1} \dots e_n^{k_n}$.

There is an obvious homomorphism of $\F_q$-algebras from $\F_q[X_1, \dots, X_n]$ to the $\F_q$-algebra ${\cal F}(q,n)$ of functions from $\F_q^n$ to $\F_q$. Considering the natural action of $S_n$ on ${\cal F}(q,n)$ one can easily see that this homomorphism is $S_n$-equivariant, i. e. it commutes with the action of $S_n$. From this it follows that $Q$ maps to the subring  ${\cal S}(q,n)$ of the symmetric functions in ${\cal F}(q,n)$. 

Since $x^q = x$ for all $x \in \F_q$, it is clear that the above homomorphism factors through the quotient:
\[\F_q\{X_1, \dots, X_n\} = \F_q[X_1, \dots, X_n] / < X_1^q-X_1, \dots, X_n^q-X_n >,\]
whose basis consists of those monomials whose exponents are smaller than $q$. So $\F_q\{X_1, \dots, X_n\}$ is finite of dimension $q^n$ and has $q^{q^n}$ elements. By using interpolation, one can deduce that the above homomorphism gives an isomorphism $\F_q\{X_1, \dots, X_n\} \simeq {\cal F}(q,n)$. This will follow by a cardinality argument. 

The ideal $I = < X_1^q-X_1, \dots, X_n^q-X_n >$ is obviously stable under the action of $S_n$. Hence $\F_q\{X_1, \dots, X_n\}$ is an $S_n$-module for the previously given action and that the isomorphism $\F_q\{X_1, \dots, X_n\} \simeq {\cal F}(q,n)$ is an isomorphism of $S_n$-modules. Under this map we see that the basis of $Q$ made of monomial symmetric functions is mapped to a basis of the symmetric function algebra ${\cal S}(q,n)$ so that the image of $Q$ in $\F_q\{x_1, \dots, x_n\}$ is isomorphic to ${\cal S}(q,n)$. 

If $E_1, \dots, E_n$ are new variables, one can also consider the quotient 
\[\F_q\{E_1, \dots, E_n\} = \F_q[E_1, \dots, E_n] / < E_1^q-E_1, \dots, E_n^q-E_n >,\]
which is again finite of dimension $q^n$ over $\F_q$. 
Consider the embedding \[\F_q[E_1, \dots, E_n] \rightarrow \F_q[X_1, \dots, X_n]\] generated by the substitutions $E_k \leadsto e_k(X_1, \dots, X_n)$. This embedding descends to a homomorphism \[\Psi : \F_q\{E_1, \dots, E_n\} \rightarrow \F_q\{X_1, \dots, X_n\} \simeq {\cal F}(q,n),\] because the embedding transports $J = < E_1^q-E_1, \dots, E_n^q-E_n >$ inside $I$.

Let ${\vec X}$ denote the tuple of variables $(X_1, \dots, X_n)$.
By diagram chasing one checks that $\Psi(\F_q\{E_1, \dots, E_n\})$ is the image of $\F_q[e_1(\vec X), \dots, e_n(\vec X)]$ under the natural projection $\F_q[X_1, \dots, X_n] \rightarrow \F_q\{X_1, \dots, X_n\}$. By standard combinatorics one knows that the dimension of the space of the symmetric functions  ${\cal S}(q,n)$ is:
\[\binom{n+q-1}{n}.\]
Summarising we have that $\Psi(\F_q\{E_1, \dots, E_n\})$ is isomorphic with the ring of symmetric functions ${\cal S}(q,n)$ and therefore the dimension of the kernel of $\Psi$ is: \[q^n - \binom{n+q-1}{n}.\]
We call this kernel ${\cal I}(q,n)$. This kernel can be seen as the set of all algebraic relations between elementary symmetric functions over a finite field. For example, if we consider the symmetric functions with two variables $e_1 = X_1 + X_2$ and $e_2 = X_1X_2$ as polynomial functions $\F_2 \rightarrow \F_2$, then:
\[e_1e_2 = (X_1+X_2)X_1X_2 = X_1^2 X_2 + X_1X_2^2 = 2X_1X_2 = 0,\]
so those functions fulfill the algebraic identity $e_1e_2 = 0$. This doesn't happens when those symmetric functions are considered over $\C$, because it is known that there the corresponding functions are algebraically independent.

The scope of this article is to show an algorithm that computes a basis of the kernel ${\cal I}(q,n)$ as $\F_q$-vector space.
The general Buchberger-M\"oller algorithm, see \cite{KR}, finds a basis of the ideal in the sense of ring theory. The algorithm presented here finds a basis as a vector space. In practical applications, if someone finds an interpolation formula for some symmetric function, and looks for a shorter definition, the basis as a vector space is more useful. Moreover, the algorithm given here for this particular problem works faster then Buchberger-M\"oller algorithm.

%%%%%%%%%%%%%%%%%%%%%%%%%%%%%%%%%%%%%%%%%%%%%%%%%%%%%%%%%%%%%%%%%%%%%%%%%%%%%%%%%%%%%%%%%%%%%%%%%%%%%%%%%%%%%%%%
%%%%%%%%%%%%%%%%%%%%%%

\section{Definitions and notations}

Consider a finite field $\F_q$ of characteristic $p$. The elements of $\F_q$ are ordered in some way, and this order is fixed.
For the rest of the paper we fix a natural number $n \geq 1$ and two sets of variables: $E_1$, $\dots$, $E_n$
and $X_1$, $\dots$, $X_n$.

\begin{definition}{\rm The set Mon$(q,n)$ is the set of all monomials $E_1^{\alpha_1}E_2^{\alpha_2}\dots E_n^{\alpha_n}$
with $0 \leq \alpha_i < q$.  There are $q^n$ many such monomials.}
\end{definition}

\begin{definition}{\rm Let $\F_q\{E_1, \dots, E_n\}$ be the vector space over $\F_q$ freely generated by Mon$(q,n)$.

$\F_q\{E_1, \dots, E_n\}$ has dimension $q^n$ over $\F_q$. 
It has also a canonical structure of finite ring induced
by the epimorphism:
\[ s : \F_q[E_1, \dots, E_n] \longrightarrow \F_q\{E_1, \dots, E_n\}\]
with Ker$(s) = (E_1^q-E_1, \dots , E_n^q-E_n)$ as ideal in $\F_q[E_1, \dots , E_n]$. 
Observe that $\F_q \models \,\, \forall x \,\, x^q=x$.}
\end{definition}

\begin{definition}{\rm For  $f:\F_q^n \rightarrow \F_q$  
and a permutation $\sigma \in S_n$ we define  $f^\sigma({\vec X}) 
= f(\sigma (\vec X))$ where $\sigma (\vec X) = (X_{\sigma(1)}, \dots, X_{\sigma(n)})$.
The function $f$ is called symmetric if  for all  $\sigma \in S_n$, $f = f^\sigma$. }
\end{definition}

\begin{definition}{\rm  Let ${\cal F}(q,n)$ denote the set of all functions
$f:\F_q^n \rightarrow \F_q$ and ${\cal S}(q,n) \subset {\cal F}(q,n)$ 
the subset of all symmetric functions. Both sets equipped with the 
point-wise operations are finite rings and finite vector spaces over $\F_q$. }
\end{definition}

\begin{definition}{\rm  For every $F \in \F_q[X_1, \dots , X_n]$ and $\sigma \in S_n$
we define $F^\sigma (\vec X) = F(\sigma(\vec X))$, where $\sigma(\vec X) = (X_{\sigma(1)}, \dots , X_{\sigma(n)})$.
$F$ is called symmetric if for all $\sigma \in S_n$, $F^\sigma = F$. }
\end{definition}

\begin{definition}{\rm For $0 \leq k \leq n$ denote by ${\cal P}^n_k$ be
the set of $k$-element subsets of $\{1, \dots, n\}$. 
Recall that the elementary symmetric polynomials $e_k(X_1, \dots , X_n)$ are defined as:
\[e_k(X_1, \dots, X_n) = \sum \limits _{ J \in {\cal P}^n_k} \prod \limits _{i \in J} X_i.\] }
\end{definition}

\begin{definition}{\rm Consider the function
\[ \Phi : \F_q \{E_1, \dots, E_n \} \longrightarrow {\cal S}(q,n) \]
defined such that
\[\forall \vec a \in \F_q^n \,\,\,\, \Phi(f)(\vec a) = f(e_1(\vec a), \dots , e_n(\vec a)), \]
where $f \in \F_q\{E_1, \dots, E_n\}$. Here we understand $E_k$ as a symbol for the elementary symmetric polynomial $e_k$.
$\Phi$ is a well defined homomorphism of finite rings and  finite $\F_q$-vector spaces. }
\end{definition}

\begin{definition}{\rm The ideal ${\cal I}(q,n) =$ Ker$(\Phi) \subset \F_q\{E_1, \dots, E_n\}$ 
is the {\bf ideal of algebraic relations} 
between elementary symmetric functions over $\F_q$.  }
\end{definition}

${\cal I}(q,n)$ is also a vector subspace of $\F_q\{e_1, \dots, e_n\}$.

We also consider the following chain of homomorphisms:

\[\F_q\{E_1, \dots, E_n\}   \xrightarrow{  \Psi  } \F_q\{X_1, \dots , X_n\} 
\xrightarrow{  \Gamma  } {\cal F}(q,n),\]

where $\Psi (P) = P(\vec e (\vec X)) $ is the substitution homomorphism 
and $\Gamma $ is the homomorphism associating to every polynomial $Q$ its polynomial function.
Of course $\Phi = \Gamma \circ \Psi$.

%%%%%%%%%%%%%%%%%%%%%%%%%%%%%%%%%%%%%%%%%%%%%%%%%%%%%%%%%%%%%%%%%%%%%%%%%%%%%%%%%%%%%%%%%%%%%%%%%%%%%%%%%%%%%%%%%
\section{The number of algebraic relations}

\begin{definition}{\rm Let WM$(q,n)$ be the set of all weakly  monotone increasing tuples $(a_1, \dots, a_n)$ 
in $\F_q$ according to the fixed order. Denote by wm$(q,n)$ the cardinality of the set WM$(q,n)$. }
\end{definition}

\begin{lemma}\label{wm}
\[{\rm dim}_{\F_q} \,\, {\cal S}(q,n) = {\rm wm}(q,n) = \binom{n+q-1}{n}.\]
\end{lemma}

{\bf Proof}:  In order to define an $f \in {\cal S}(q,n)$, 
it is enough to define its values for every ${\vec \iota} \in$ WM$(q,n)$.
This number of sequences is the same as the number of possibilities to put $n$ unsigned balls in $q$ numbered urns, see \cite{WF}. As done there, the $q$ urns can be represented by $q+1$ vertical bars and the $n$ balls as $n$ circles. A possible distribution as it follows:
\[| \circ | | \circ \circ ||\]
Because the first and the last symbols in distribution must be bars, we have to distribute $n$ circles in $n+q-1$ positions and to complete the remaining positions with bars. There are
\[\binom{n+q-1}{n}\]
possibilities to do so.\qed

\begin{definition}{\rm Consider the following matrix 
$M(q,n) \in {\cal M}({\rm wm}(q,n) \times q^n , \F_q)$. The rows of $M(q,n)$ are 
indexed using the tuples $\vec \iota \in $ WM$(q,n)$, 
the columns are indexed using the monomials $m \in $ Mon$(q,n)$, 
and if $M(q,n) = (a(\vec \iota, m)\,\,|\,\, \vec \iota \in {\rm WM}(q,n), m \in {\rm Mon}(q,n))$,
\[a(\vec \iota, m) = [\Phi(m)](\vec \iota) = e_1^{\alpha_1}({\vec \iota}) \dots e_n^{\alpha_n}({\vec \iota}),\]
where $m = E_1^{\alpha_1}\dots E_n^{\alpha_n}$.
 }
\end{definition}

\begin{lemma}\label{surj}

The morphism $\Phi: \F_q \{E_1, \dots, E_n \} \rightarrow {\cal S}(q,n)$ with 
${\rm Ker}\,\,\Phi = {\cal I}(q,n)$ 
is surjective.

\end{lemma}

{\bf Proof}: The proof consists of two steps. 

{\bf Step} 1: Let $f : \F_q^n \rightarrow \F_q $ be some function, for the moment not necessarily symmetric. 
For $a \in \F_q$ 
define the polynomial $h_a \in \F_q[X]$:
\[h_a(X) = \prod \limits _{\lambda \in \F_q \setminus \{a\}} \frac{X-\lambda}{a - \lambda}.\]
Observe that $h_a(a) = 1$ and $h_a(\F_q\setminus \{a\}) = 0$. For a tuple $\vec a \in \F_q^n$ 
define $h_{\vec a} \in \F_q[X_1, \dots, X_n]$:
\[h_{\vec a}(\vec X) = h_{a_1}(X_1)\dots h_{a_n}(X_n).\]
A polynomial interpolating $f$ is:
\[H(\vec X) = \sum \limits _{\vec a \in \F_q^n} h_{\vec a}(\vec X) f(\vec a).\]
We observe that for all $\sigma \in S_n$, 
${(h_{\vec a})} ^\sigma = h_{\sigma ^ {-1}(\vec a)}$. 
If the function $f$ is symmetric, then $f^\sigma = f$ and it follows:
\[H^\sigma(\vec X) = \sum \limits _{\vec a \in \F_q^n} h_{\vec a}^\sigma(\vec X) f(\vec a) 
= \sum \limits _{\sigma ^{-1}(\vec a) \in \F_q^n} h_{\sigma ^{-1}(\vec a)}(\vec X) f(\sigma ^{-1}(\vec a)) = H(\vec X).\]
We proved that the interpolation algorithm applied to a symmetric function leads to a symmetric polynomial. 
Observe also that all exponents occurring in $H$ are $< q$.

{\bf Step} 2: We repeat the argument that a symmetric polynomial 
is a polynomial in elementary symmetric polynomials as given in \cite{BLW} and reformulated in \cite{BS}.
The following total order is defined over the set of monomials in $\vec X$: 
$X_1^{\alpha_1} \dots  X_n^{\alpha_n} < X_1^{\beta_1} \dots  X_n^{\beta_n}$ 
if and only if $\sum \alpha_i < \sum \beta_i$ or 
$\sum \alpha_i = \sum \beta_i$ but $(\alpha_i) < (\beta_i)$ lexicographically. This is the graded lexicographic order.
For a polynomial $H(\vec X) \in \F_q[\vec X]$ define
Init$(H)$ to be the maximal monomial occurring in $H$, according to this order. 
It follows from symmetry that Init$(H)$ has the form 
$cX_1^{\gamma_1} \dots X_n^{\gamma_n}$ with $\gamma_1 \geq \gamma_2 \geq \dots \geq \gamma_n$ 
with $c \in \F_q \setminus\{0\}$. 
Consider the polynomial:
\[ H_1(\vec X) = H(\vec X) - c e_1(\vec X)^{\gamma_1-\gamma_2}e_2(\vec X)^{\gamma_2 - \gamma_3} 
\dots e_{n-1}(\vec X)^{\gamma_{n-1}-\gamma_n} e_n(\vec X)^{\gamma_n}.\]
Observe that Init$(H_1)$ $<$ Init$(H)$. Continue by constructing in the same way a polynomial
$H_2$ with Init$(H_2)$ $< $ Init$(H_1)$, and so on. This process ends in finitely many steps. 
Adding the $S_i$-monomials defined during the process,
one gets a polynomial $F \in \F_q[e_1, \dots , e_n]$ with the property that for all $\vec a \in \F_q^n$, 
$F(e_1(\vec a), \dots , e_n(\vec a)) = f(\vec a)$. Finally, 
observe that all exponents occurring in $F$ are again $< q$, so 
$F \in \F_q\{e_1, \dots , e_n\}$ and $\Phi(F) = f$. 

\qed

\begin{theo} \label{main}
The rank of the matrix $M(q,n)$ is maximal:
\[{\rm rank} \,\, M(q,n) = {\rm wm}(q,n) = \binom{n+q-1}{n}.\]
The dimension of the ideal ${\cal I}(q,n)$ of algebraic relations as a vector space over $\F_q$ is:
\[ {\rm dim}_{\F_q} {\cal I}(q,n) = q^n - {\rm wm}(q,n) = q^n - \binom{n+q-1}{n}.\]

\end{theo}

{\bf Proof}: Follows directly from Lemma \ref{wm} and the  Lemma \ref{surj}. \qed

\begin{remark}{\rm
If $q$ is kept constant and $n \rightarrow \infty$,
\[ \frac{{\rm dim}_{\F_q} {\cal I}(q,n)}{{\rm dim}_{\F_q} \F_q\{S_1, \dots , S_n\} } \,\, \longrightarrow \,\, 1.\]
If $n$ is kept constant and $q \rightarrow \infty$,
\[ \frac{{\rm dim}_{\F_q} {\cal I}(q,n)}{{\rm dim}_{\F_q} \F_q\{S_1, \dots , S_n\} } \,\, \longrightarrow \,\, 1 - 
\frac{1}{n!}.\]}
\end{remark}

Indeed, if $q$ is kept constant and $n \rightarrow \infty$,
\[\frac{q^n - \binom{n+q-1}{n}}{q^n} = 1-\frac{1}{(q-1)!}\frac{(n+1)\dots (n+q-1)}{q^n} \,\, \longrightarrow \,\,1.\]
If $n$ is kept constant and $q \rightarrow \infty$,
\[\frac{q^n - \binom{n+q-1}{n}}{q^n} =  1 - \frac{1}{n!} \frac{q(q+1)\dots(q+n-1)}{q^n} \,\, \longrightarrow \,\, 1 - 
\frac{1}{n!}.\]

\begin{remark}{\rm Define the set of non-monotone tuples NM$(q,p)$ as the set of tuples 
$(a_1, \dots , a_n)$ with $a_i \in \F_q$
such that there are $1 \leq i < j \leq n$ with $a_i > a_j$ according to the order fixed on $\F_q$. 
Let nm$(q,n)$ be the cardinality of the set NM$(q,n)$.
According to Theorem \ref{main}, ${\cal I}(q,n)$ has dimension nm$(q,n)$. But is there 
any natural correspondence between NM$(q,n)$ and a basis of the vector space ${\cal I}(q,n)$? As far I know, this problem is open. }
\end{remark}

\begin{remark}{\rm Recall the chain of homomorphisms:

\[\F_q\{E_1, \dots, E_n\}   \xrightarrow{  \Psi  } \F_q\{X_1, \dots , X_n\} 
\xrightarrow{  \Gamma  } {\cal F}(q,n),\]

where $\Psi (P) = P(\vec e (\vec X)) $ is the substitution homomorphism 
and $\Gamma $ is the homomorphism associating to every polynomial $Q$ its polynomial function, and $\Phi = \Gamma \circ \Psi$. Using the interpolation part of the proof of Lemma \ref{surj} one sees that $\Gamma $ is an isomorphism of rings and vector spaces. 
Indeed, $\Gamma$ is a surjective homomorphism and both rings have $q^{q^n}$ elements. }
\end{remark}

\begin{cor}\label{kerpsi}
${\rm Ker} \, \Psi = {\cal I}(q,n)$ and ${\rm Im} \, \Psi = \Gamma ^{-1} ({\cal S}(q,n))$. Consequently, the 
subring of symmetric polynomials in $\F_q\{X_1, \dots , X_n\}$ is a vector space of dimension {\rm wm}$(q,n)$ over $\F_q$.
\end{cor}

%%%%%%%%%%%%%%%%%%%%%%%%%%%%%%%%%%%%%%%%%%%%%%%%%%%%%%%%%%%%%%%%%%%%%%%%%%%%%%%%%%%%%%%%%%%%%%%%%%%%%%%%%
%%%%%%%%%%%%%%%%%%%%%%%%%%%%%%%%

\section{Deduction procedure}

Before presenting the specific algorithm that generates a basis of the ideal of algebraic identities, I briefly recall the Gauss Algorithm over some field $K$.

{\bf Gauss Algorithm}:  Consider the extended matrix $M$ of a system of $e$ many linear equations in $v$ unknowns. The matrix has $e$ rows and $v+1$ columns. The algorithm finds out if the system is solvable. It also finds out on how many parameters the solution depends and generates a parametric solution of the system. 

At the beginning one completes an array $\pi$ with the values $\pi(0) = 0$, $\dots$, $\pi(v-1) = v-1$. This array will accumulate the total permutation of lines, produced by the algorithm. 

Let $r$ be a variable in  which the algorithm computes the rank of the restricted matrix of the system. Let $m = \min(e, v)$. The Gauss Algorithm is a repetition of Gauss Steps according to a variable $s$ running from $0$ to $m-1$ inclusively. If the Gauss Step returns {\it false} at step $s$, the loop is broken. If not, $r$ is increased by $1$ and the corresponding Gauss Step is performed. 

The Gauss Step of $s$ works as follows. In every column $c$ with $s \leq c < v$ one looks for a row $w$ with $s \leq w < e$ such that $M(w, c) \neq 0$. If no such an element is found, the Gauss Step returns {\it false} and stops.  Once such an element is found: if $c \neq s$, the columns $c$ and $s$ are inter-changed and also, the values of $\pi(s)$ and $\pi(c)$ are inter-changed. Further, if $s \neq w$, the rows $w$ and $s$ are inter-changed. Let now $a = M(s,s)$. Then the row $s$ is divided by $a$. Further, from all the rows $d$ with $d > s$ is subtracted the row $s$ times a field value $b$, where $b = M(d, r)$. Because of this, all elements of the column $s$ which are below $M(s,s)$ become $0$. Finally, the Gauss Step returns {\it true}. 

The decision on the number of solutions is made as follows. If the restricted rank $r$ is equal with the number of unknowns $v$, there is only one solution. If $r < e$, there are two possibilities: (1) there are elements $M(v, i) \neq 0$ with $i \geq r$. In this case there are no solutions. Or (2), there are no such elements. Then the unknowns which corresponds now, after all permutations of columns (unknowns), to the columns $r$, $r+1$, $\dots$, $v$, are independent parameters. The other unknowns are computed successively, started to the unknown corresponding to column $v-1$ after the permutation, continuing to the unknown corresponding to the column $v-2$, and finally ending with the unknown corresponding to the column $0$. 

\qed

The following algorithm is able to find a basis over $\F_q$ for the vector space ${\cal I}(q,n)$ of
algebraic relations between elementary symmetric functions
over $\F_q$.

\begin{enumerate}

\item Consider $q^n$ many new unknowns $Y_m$ indexed using the set Mon$(q,n)$, 
and the following homogenous system $\Sigma$ of wm$(q,n)$ many linear equations 
indexed using the set WM$(q,n)$:
\[(\vec \iota): \,\,\,\,\,\,\,\, 
\sum \limits _{ m \in {\rm Mon}(q,n)}[\Phi(m)](\vec \iota)Y_m = 0.\]
The matrix of this  linear homogenous system is the matrix $M(q,n)$ defined in the previous section.
One sees that for any polynomial $P \in \F_q\{e_1, \dots ,e_n\}$ following holds:
\[P(\vec e) = \sum \limits _{ m \in {\rm Mon}(q,n)} y_m \,\,\,m(\vec e) \,\,\,\,\in 
\,\,{\cal I}(q,n) \,\, \Leftrightarrow \,\,(y_m) \in {(\F_q)}^{q^n} \,\, {\rm satisfies}\,\,\Sigma.\]

\item Using Gauss' Algorithm over $\F_q$ transform $M(q,n)$ in an upper triangular matrix. Recall that $M(q,n)$ has
maximal rank equal with ${\rm wm}(q,n)$.

\item Introduce a tuple $\vec t$ of $q^n - {\rm wm}(q,n)$ many new parameters and  
compute the parametric solution of $\Sigma$, consisting of linear functions in $\vec t$:
\[(Y_m(\vec t))_{m \in {\rm Mon}(q,n)}.\]

\item Using the equivalence from (1) one has that:
\[{\cal I}(q,n) = \Big \{ \sum \limits _{ m \in {\rm Mon}(q,n)} Y_m(\vec t)\,\,\, m\,\,\,\,| \,\, 
\vec t \in (\F_q)^{q^n - {\rm wm}(q,n)} \,\,\Big \}.\] 
For $i = 1, \dots, q^n-{\rm wm}(q,n)$ set the parameter tuple $\vec t_i = (0,0, \dots, t_i = 1, 0,\dots , 0)$ in the 
general form \[P_{\vec t} = \sum \limits _{ m \in {\rm Mon}(q,n)} Y_m(\vec t) m\]
 and call the result of the substitution $P_i=P_{\vec t_i}$. The mapping
$\vec t \leadsto P_{\vec t}$  is an isomorphism of vector spaces over $\F_q$ because it is a surjective linear mapping
between vector spaces of equal dimensions. As an isomorphism, this mapping transports basis to basis, 
so $\{\,P_i\,|\,i = 1, \dots, q^n-{\rm wm}(q,n)\,\}$ is a basis of the vector space ${\cal I}(q,n)$ over $\F_q$.

\end{enumerate}

The fact that this algorithm finds a basis of the ideal of identities of symmetric polynomials follows directly from the theory developed above and from its description. It remains to show only that the coefficients of the basis polynomials always belong to the base field. This shall be shown in the next Section. Here some examples will be given.

{\bf Example}. $\F_2[X_1, X_2]$: dim$_{\F_2} {\cal I}(2,2) = 2^2 -$ wn$(2,2) = 4 - 3 = 1$. The monomial basis of the vector space $\F_2\{E_1, E_2\}$ is the set $\{1, E_1, E_2, E_1E_2\}$. The matrix $M(2,2)$ is the following:
\begin{center}
\begin{tabular}{ c| c c c c}
\cell{} & \cell{1} & \cell{$e_1$} & \cell{$e_2$} & \cell{$e_1e_2$}  \\
\hline
\cell{00} & \cell{1} & \cell{0} & \cell{0} & \cell{0}  \\ 
\cell{01} & \cell{1} & \cell{1} & \cell{0} & \cell{0}  \\ 
\cell{11} & \cell{1} & \cell{0} & \cell{1} & \cell{0}
\end{tabular}
\end{center}
Linear variables $\{Y_1, Y_2, Y_3, Y_4\}$ are associated to the columns. The linear system of equations over $\F_2$:
\begin{eqnarray*}
Y_1 &=&0,\\
Y_1+Y_2 &=& 0,\\
Y_1+Y_3 &=& 0,
\end{eqnarray*}
has the parametric solution $(Y_1, Y_2, Y_3, Y_4) = (0,0,0,t)$, where $t \in \F_2$ is a parameter. It follows that:
\[{\cal I}(2,2) = \{t e_1e_2\,|\,t \in \F_2\} = \{0, e_1e_2\}.\]
The only one nontrivial algebraic relation  between elementary symmetric functions is in this case the following:
\begin{eqnarray*}
e_1e_2 & = & 0.
\end{eqnarray*}
We verified in the introduction that $e_1e_2=0$ as a function.

\vspace{3mm}

{\bf Example}.  $\F_2[X_1, X_2, X_3]$:  dim$_{\F_2} {\cal I}(2,3) = 2^3 -$ wn$(2,3) = 8 - 4 = 4$.  The monomial basis of the vector space $\F_2\{E_1, E_2, E_3\}$ is the set $\{1, E_1, E_2, E_3, E_1E_2, E_1E_3, E_2E_3, E_1E_2E_3\}$. The matrix $M(2,3)$ is the following:
\begin{center}
\begin{tabular}{ c| c c c c c c c c}
\cell{} & \cell{1} & \cell{$e_1$} & \cell{$e_2$} & \cell{$e_3$} & \cell{$e_1e_2$} & \cell{$e_1e_3$} & \cell{$e_2e_3$} & \cell{$\,\,\,\,\,\,e_1e_2e_3$}  \\
\hline
\cell{000} & \cell{1} & \cell{0} & \cell{0} & \cell{0} & \cell{0} & \cell{0}& \cell{0}& \cell{0}  \\ 
\cell{001} & \cell{1} & \cell{1} & \cell{0} & \cell{0} & \cell{0}& \cell{0}& \cell{0}& \cell{0} \\ 
\cell{011} & \cell{1} & \cell{0} & \cell{1} & \cell{0} & \cell{0}& \cell{0}& \cell{0}& \cell{0} \\
\cell{111} & \cell{1} & \cell{1}& \cell{1}& \cell{1}& \cell{1}& \cell{1}& \cell{1}& \cell{1}
\end{tabular}
\end{center}
Gauss' Algorithm produces the following matrix:
\begin{center}
\begin{tabular}{c c c c c c c c c}
\cell{} & \cell{1} & \cell{0} & \cell{0} & \cell{0} & \cell{0} & \cell{0} & \cell{0} & \cell{0}  \\ 
\cell{} &\cell{0} & \cell{1} & \cell{0} & \cell{0} & \cell{0}  & \cell{0} & \cell{0} & \cell{0} \\ 
\cell{} &\cell{0} & \cell{0} & \cell{1} & \cell{0} & \cell{0} & \cell{0} & \cell{0} & \cell{0} \\
\cell{} &\cell{0} & \cell{0} & \cell{0} & \cell{1} & \cell{1} & \cell{1} & \cell{1} & \cell{1}
\end{tabular}
\end{center}
Linear variables $\{Y_1, Y_2, Y_3, Y_4, Y_5, Y_6, Y_7, Y_8\}$ are associated to the columns. 
The linear system of equations over $\F_2$
has the parametric solution: \[(Y_1, Y_2, Y_3, Y_4, Y_5, Y_6, Y_7, Y_8) = (0,0,0,t_1 + t_2 + t_3 + t_4, t_1, t_2, t_3, t_4 ),\] where $t_1, t_2, t_3, t_4 \in \F_2$ are parameters. It follows that:
\[{\cal I}(2,3) = \{(t_1 + t_2 + t_3 + t_4) e_3 + t_1 e_1e_2 + t_2 e_1e_3 + t_3 e_2e_3 + t_4 e_1e_2e_3
\,|\,t_1, t_2, t_3, t_4 \in \F_2\}.\]
By replacing the parameter vector $(t_1, t_2,t_3,t_4)$ with the unit vectors $(1,0,0,0)$, $\dots$, $(0,0,0,1)$, following basis of linear independent algebraic relations is found:

\begin{eqnarray*}
e_3 + e_2e_3 & = & 0 \\
e_3 + e_1e_3 & = & 0 \\
e_3 + e_1e_2 & = & 0 \\
e_3 + e_1e_2e_3 & = & 0
\end{eqnarray*}

\begin{remark}\rm In this algorithm, we can skip the line corresponding to the tuple $(0,0, \dots, 0)$ and the column corresponding to the monomial $1$. This does not change the result.
\end{remark}

\begin{remark}\rm If $n = 1$, the monomials are $\{1, X, X^2, \dots, X^{q-1}\}$ and the tuples are the $q$ elements of $\F_q$. It follows that the matrix $M(q,1)$ is nothing but the Vandermonde matrix over the field $\F_q$. The ideal ${\cal I}(q,1)$ is always $0$.
\end{remark}

%%%%%%%%%%%%%%%%%%%%%%%%%%%%%%%%%%%%%%%%%%%%%%%%%%%%%%%%

\section{Coefficients in $\F_p$}

\begin{theo}\label{coeffp}
Let $\F_q$ be some finite field, $p = {\rm char} \,\, \F_q$ and $n\geq 2$. Then the algorithm presented above finds
a basis of ${\cal I}(q,n)$ consisting of polynomials with coefficients in the prime field $\F_p$. In particular, such a basis
always exists.
\end{theo}

Before proving the Theorem \ref{coeffp}, we must fix some notations concerning Gauss' Algorithm.

{\bf Definition}: For  $i < j$ we define:

$A(i, j)$ means that the equation $i$ multiplied with an apropriate element is added to equation $j$. The element is 
chosen such that the first non-zero coefficient in the equation $j$ becomes $0$.

$L(i,j)$ means that equations (lines) $i$ and $j$ are inter-changed. 

$C(i, j)$ means that the columns $i$ and $j$ are inter-changed. 

{\bf Definition}: Here is the step number $i$ of the deterministic Gauss' Algorithm.
If $a_{i,i} = 0$ and the whole line $i$ consists only of zeros, find the first $j > i$ such that the line $j$ contains
non-zero elements, and apply $L(i,j)$. If $a_{i,i}$ continues to be zero, find the first $k > i$ such that $a_{i,k} \neq 0$
and apply $C(i,k)$. Now, for each $ r > i$; if the element $a_{r,i} \neq 0 $ below $a_{i,i}$, apply $A(i,r)$.

{\bf Definition}: Let $K$ be some field and $S$, $S'$ two systems of linear equations over $K$; both of them consisting
of $e$ equations with $u$ unknowns. We say that Gauss' Algorithm works in parallel for systems $S$ and $S'$ if the
application of Gauss' Algorithm on the systems leads to the same sequence of operations $O_1, O_2, \dots , O_s$ in the above
notation. (For example, $O_1 = A(1,2)$, .  .  . , $O_{s} = L(3,4)$, $O_{s+1} = C(3,5)$, .  .  .  and so on.)

\begin{lemma}\label{permutsys}
Let $K$ be a field and $S$, $S'$ two homogenous systems of linear equations over $K$, satisfying the 
following conditions:
\begin{enumerate}
\item  $S$ and $S'$ have both $e$ equations and
$u$ unknowns, $k = u-e \geq 0$ and  ${\rm rank}\,\,S = e$.
\item $S'$ has been obtained from $S$ by some permutation of lines (equations). 
\item Gauss' Algorithm works in parallel over $S$ and $S'$.
\end{enumerate}
In this situation, Gauss' Algorithm independently applied for the systems $S$ and $S'$ 
computes the same parametrization of the common space of solutions:
\[\forall \,\, i = 1, \dots, u\,\,\,\,\,\,\,\,\,\, Y_i(\vec t) = Y_i'(\vec t),\]
where ${\vec t} = (t_1, t_2, \dots, t_k)$ is the tuple of parameters. 
\end{lemma}

{\bf Proof}: Let $M {\vec Y} = {\vec 0}$ be the system $S$ of linear equations as in the statement. Running Gauss' Algorithm forth and back, we find that columns with indexes $i_1, \dots, i_e$ build a non-singular $e \times e$ minor $A$. Let $j_1, \dots j_k$ be the indexes of the left columns. For $s = 1, \dots, k$, we set $Y_{j_s} = t_s$, where $t_1, \dots, t_s$ are independent  parameters. 

%Let $y_i$ be the vector $(y_{i_1}, \dots, y_{i_e})$.
The parametric solution is obtained from:
\[A \begin{pmatrix} Y_{i_1} \\ \vdots \\ Y_{i_e} \end{pmatrix} = - \sum \limits _{s=1} ^k {\vec c}_{j_s} t_s,\]
where ${\vec c}_{j_s}$ are the column of $M$ left outside $A$. The parametric solution has the form:
\[\begin{pmatrix} Y_{i_1} \\ \vdots \\ Y_{i_e} \end{pmatrix} = \sum \limits _{s=1} ^k {\vec d}_{j_s} t_s,\]
completed with the $k$ many $Y_{j_s} = t_s$ already done. Here the column vector $ {\vec d}_{j_s}$ is the unique solution of the system of linear equations:
\[A \begin{pmatrix} Y_{i_1} \\ \vdots \\ Y_{i_e} \end{pmatrix} = - {\vec c_{j_s}}.\]

Now consider the system $S'$ of  linear equations  $M' {\vec Y} = {\vec 0}$. As Gauss' Algorithm works in parallel for the systems $S$ and $S'$, by running it forth and back, we find the same columns $i_1, \dots , i_e$ to build a non-singular minor $A'$. But the matrix $M'$ consists of the same lines as $M$ in a permuted order, and the same permutation is used to get $A'$ from $A$ and $\vec {c_{j_s}'}$ from $\vec {c_{j_s}}$. The parametric solution is again:
\[\begin{pmatrix} Y_{i_1} \\ \vdots \\ Y_{i_e} \end{pmatrix} = \sum \limits _{s=1} ^k {\vec d}_{j_s} t_s,\]
completed with the $k$ many $Y_{j_s} = t_s$, where  the column vector $ {\vec d}_{j_s}$ is the unique solution of the system of linear equations:
\[A' \begin{pmatrix} Y_{i_1} \\ \vdots \\ Y_{i_e} \end{pmatrix} = - \vec {c_{j_s}'}.\]
This system is the same system as previous one, up to the order in which the equations are displayed, and we know that permuting the equation in a system with unique solution, does not change this solution. 
\qed

%%%%%%%%%%%%%%%%%%%%%%%%%%%%%%%%%%%%%%%%%%%%%%%%%%%%%%%

{\bf Proof of the Theorem \ref{coeffp}}: If $\F_q = \F_p$ there is nothing to prove. 
Consider some automorphism $\varphi \in {\rm Gal}(\F_q / \F_p)$. The fact that $\varphi$ is a power of Frobenius'
Automorphism is not relevant here. Consider the system $\Sigma$ used in the algorithm given above and the 
system $\Sigma' = \varphi(\Sigma)$.

{\it Claim 1: $\Sigma'$ can be obtained from $\Sigma$ by some permutation of equations.} Indeed, the elements of the matrix 
$\varphi(M(q,n))$ are $\varphi(a(\vec \iota, m)) = \varphi([\Phi(m)](\vec \iota)) = [\Phi(m)](\vec {\varphi(\iota)})$.
The automorphism $\varphi$ is a permutation of $\F_q$.
Consequently, the line formerly indexed $\vec \iota$ shall be found in $\Sigma'$ at the index obtained by the weakly monotone
reordering of the tuple ${\varphi(\vec \iota)}$. As the coefficients of the system are values of symmetric functions for this tuple, the order inside the tuple does not matter.

{\it Claim 2: Gauss' Algorithm works in parallel over $\Sigma$ and $\Sigma'$.} The
coefficients in $\Sigma'$ are images
of corresponding coefficients in $\Sigma$ by $\varphi$. 
This situation remains true after every computation step done by Gauss Algorithm.
In particular, at every step one has in both systems $\Sigma$ and $\Sigma'$ 
the same situation concerning coefficients (matrix entries) which are zero or not. So after every step, the same
decision concerning the next step shall be taken: an $A(i,j)$ or a $C(i,j)$. 

So all conditions requested by Lemma \ref{permutsys} are satisfied, and Gauss' Algorithm computes the same
parametrization $(Y_i(\vec t))$ for both systems $\Sigma$ and $\varphi(\Sigma)$. 
So for all $i$, $Y_i(\vec t) = \varphi(Y_i(\vec t))$; 
and this takes place for all automorphisms 
$\varphi \in {\rm Gal}(\F_q / \F_p)$. It follows that the linear 
functions $Y_i(\vec t)$ have coefficients in $\F_p$, and
the same is true for the basis of ${\cal I}(q,n)$ found by the algorithm. \qed

%%%%%%%%%%%%%%%%%%%%%%%%%%%%%%%%%%%%%%%%%%%%%%%%%%%%%%%%%%%%%%%%%%%%%%%%%%%%%%%%%%%%%%%%%%%%%%%%%%%%%%%%%%%%%%%%%%%%

\section{Further examples}

 This algorithm has been implemented by the author using the language C$\#$ on Visual Studio, and ran for the fields $\F_q$ 
 with $q \in\{2,3,4,5,7,8,9,11, 16, 25, 27, 49, 81 \}$ and values of $n \leq 5$.
 The implementation uses the fact that the elementary symmetric polynomials in $x_1, \dots, x_n$
 can be computed all at a time in quadratic time and linear space.  We show here only some  examples. 

$\F_3[x_1, x_2]$. dim$_{\F_3} {\cal I}(3,2) = 3^2-$ wn$(3,2) = 9 - 6 = 3$. Following basis has been found:

\begin{eqnarray*}
 2 e_1 e_2 +  e_1 e_2^2 & = & 0 \\
 e_2 +  e_2^2 +  e_1^2 e_2 & = & 0\\
 e_2 +  e_2^2 +  e_1^2 e_2^2 & = & 0
\end{eqnarray*}

\vspace{3mm}

$\F_3[x_1, x_2, x_3]$.  dim$_{\F_3} {\cal I}(3,3) = 3^3 -$ wn$(3,3) = 27 - 10 = 17$.
Following basis has been found:

\begin{eqnarray*}
2 e_2 e_3^2 +  e_1 e_3 & = & 0 \\
2 e_2 e_3 +  e_1 e_3^2 & = & 0 \\
  e_2 e_3^2 +  e_2^2 e_3^2 & = & 0 \\
  e_2 e_3^2 +  e_1 e_2 e_3 & = & 0 \\
  e_2 e_3 +  e_1 e_2 e_3^2 & = & 0 \\
  e_2 e_3 + 2 e_1 e_2 +  e_1 e_2^2 & = & 0 \\ 
 2 e_2 e_3^2 +  e_1 e_2^2 e_3 & = & 0 \\
 2 e_2 e_3 +  e_1 e_2^2 e_3^2 & = & 0 \\
  e_2 e_3 +  e_2^2 e_3 & = & 0 
\end{eqnarray*}
\begin{eqnarray*}
  e_2 e_3 +  e_1^2 e_3 & = & 0 \\
  e_2 e_3^2 +  e_1^2 e_3^2 & = & 0 \\
  e_2 + 2 e_2 e_3^2 +  e_2^2 +  e_1^2 e_2 & = & 0 \\ 
 2 e_2 e_3 +  e_1^2 e_2 e_3 & = & 0 \\
 2 e_2 e_3^2 +  e_1^2 e_2 e_3^2 & = & 0 \\ 
  e_2 +  e_2 e_3^2 +  e_2^2 +  e_1^2 e_2^2 & = & 0 \\
  e_2 e_3 +  e_1^2 e_2^2 e_3 & = & 0 \\
  e_2 e_3^2 +  e_1^2 e_2^2 e_3^2 & = & 0 
\end{eqnarray*}

\vspace{3mm}

$\F_4[x_1, x_2]$: The field $\F_4$ has been realized using the polynomial $X^2+X+1$ over $\F_2$.
dim$_{\F_4} {\cal I}(4,2) = 4^2 -$ wn$(4,2) = 16 - 10 = 6$. 
The found basis has coefficients in $\F_2$:

\begin{eqnarray*}
 e_1 e_2 +  e_1^2 e_2^2 & = & 0 \\
 e_1 e_2^2 +  e_1^2 e_2^3 & = & 0 \\
 e_1 e_2^3 +  e_1^2 e_2 & = & 0 \\
 e_1 e_2^2 +  e_1^3 e_2 & = & 0 \\
 e_1 e_2^3 +  e_1^3 e_2^2 & = & 0 \\
 e_1 e_2 +  e_1^3 e_2^3 & = & 0 
\end{eqnarray*}

\vspace{3mm}

$\F_5[x_1, \dots, x_5]$: dim$_{\F_5} {\cal I}(5,5) = 5^5 -$ wn$(5,5) = 3125 - 126 = 2999$.
The first identity found was:

\begin{eqnarray*}
 4 e_4^2 e_5 +  e_4^4 e_5 + 3 e_3 e_4 e_5^2 + 4 e_3 e_4^3 e_5^2 + 3
e_3 e_4^4 e_5^2 + 4 e_3^2 e_4 e_5^3 +  e_3^2 e_4^4 e_5^3 = 0.
\end{eqnarray*}

\vspace{3mm}

$\F_7[x_1, \dots , x_4]$:  dim$_{\F_7} {\cal I}(7,4) = 7^4 -$ wn$(7,4) = 2401 - 210 = 2191$.
The first identity found was:

\[  e_4 + 6 e_4^4 + 6 e_3^6 e_4 +  e_3^6 e_4^4 + 5 e_2 e_4^2 + 2
e_2 e_4^5 + 5 e_2 e_3^2 e_4^2 + 2 e_2 e_3^2 e_4^5 + 5 e_2 e_3^4 e_4^2 + \]
\[+ 2 e_2 e_3^4 e_4^5 + 6 e_2^2 e_4^3 +  e_2^2 e_4^6 + 6 e_2^2 e_3^2
e_4^3 +  e_2^2 e_3^2 e_4^6 + 6 e_2^2 e_3^4 e_4^3 +  e_2^2 e_3^4 e_4^6 = 0.\]

%%%%%%%%%%%%%%%%%%%%%%%%%%%%%%%%%%%%%%%%%%%%%%%%%%%%%%%%%%%%%%%%%%%%%%%%%%%%%%%%%%%%%%%%%%%%%%%%%%%%%%%%%%%%%%%%%%

% \bibliographystyle{eptcs}
% \bibliography{biblio}

\end{document}